\let\ifarxiv=\iftrue
\documentclass[a6paper,12pt]{article}
\ifarxiv\pdfoutput=1\fi
\usepackage{amsmath,amssymb}

\ifnum\pdfoutput=1
\usepackage[bookmarks=true,hyperfigures=true]{hyperref}
\usepackage[final]{graphicx}
\else
\usepackage[bookmarks=true,hyperfigures=true,hypertex]{hyperref}
\usepackage[draft]{graphicx}
\fi

\usepackage[nosort]{cite}

\ifx\hypersetup\sadfkjashdfkxja\else
\hypersetup{pdftitle={Mass Deformation of Super Yang-Mills in D=2+1}}
\hypersetup{pdfauthor={Abhishek Agarwal}}
\hypersetup{pdfsubject={}}
\hypersetup{pdfkeywords={Mass-Deformation, Super-Poincare}}
\hypersetup{plainpages=false}
\hypersetup{pdfpagemode=none}
\hypersetup{pdfpagelabels=true}
\hypersetup{bookmarksnumbered=true}
\hypersetup{pdfstartview=FitH}
\hypersetup{pdfpagemode=UseNone}
\hypersetup{pdfpagemode=UseOutlines}
\hypersetup{colorlinks=false}
\hypersetup{citebordercolor={.5 1 .5}}
\hypersetup{urlbordercolor={.5 1 1}}
\hypersetup{linkbordercolor={1 .7 .7}}
\hypersetup{pdfborder={0 0 1 [3]}}
\fi

\numberwithin{equation}{section}

\usepackage[a4paper,left=2.8cm,right=2.8cm,top=2.8cm,bottom=2.8cm]{geometry}

\allowdisplaybreaks[3]

\usepackage[font={small},labelfont={bf},width=0.85\textwidth]{caption}

\makeatletter
\let\old@startsection=\@startsection
\renewcommand{\@startsection}[6]{\old@startsection{#1}{#2}{#3}{#4}{#5}{#6\mathversion{bold}}}
\makeatother

\def\[{\begin{equation}}
\def\]{\end{equation}}
\def\<{\begin{eqnarray}}
\def\>{\end{eqnarray}}

\newcommand{\nln}{\nonumber\\}

\newcommand{\earel}[1]{\mathrel{}&\hspace{-2\arraycolsep}#1\hspace{-2\arraycolsep}&\mathrel{}}
\newcommand{\eq}{\earel{=}}

\ifx\genfrac\sdflkaj

\else

\fi

\newcommand{\tr}{\mbox{tr}}

\let\oldPhi=\Phi
\let\oldPsi=\Psi
\let\oldGamma=\Gamma
\let\oldDelta=\Delta
\let\oldSigma=\Sigma
\let\oldLambda=\Lambda
\let\oldTheta=\Theta
\let\oldPi=\Pi
\renewcommand{\Phi}{\mathnormal{\oldPhi}}
\renewcommand{\Psi}{\mathnormal{\oldPsi}}
\renewcommand{\Gamma}{\mathnormal{\oldGamma}}
\renewcommand{\Sigma}{\mathnormal{\oldSigma}}
\renewcommand{\Delta}{\mathnormal{\oldDelta}}
\renewcommand{\Theta}{\mathnormal{\oldTheta}}
\renewcommand{\Lambda}{\mathnormal{\oldLambda}}
\renewcommand{\Pi}{\mathnormal{\oldPi}}

\newcommand{\matr}[2]{\left(\begin{array}{#1}#2\end{array}\right)}

\ifx\href\asklfhas\newcommand{\href}[2]{#2}\fi

\makeatletter
\def\mr@ignsp#1 {\ifx\:#1\@empty\else #1\expandafter\mr@ignsp\fi}%
\newcommand{\multiref}[1]{\begingroup%\let\protect\string%
\xdef\mr@no@sparg{\expandafter\mr@ignsp#1 \: }%
\def\mr@comma{}%
\@for\mr@refs:=\mr@no@sparg\do{\mr@comma\def\mr@comma{,}\ref{\mr@refs}}%
\endgroup}
\makeatother

\newcommand{\hypref}[2]{\ifx\href\asklfhas #2\else\href{#1}{#2}\fi}

\renewcommand{\eqref}[1]{(\multiref{#1})}

\makeatletter
\newlength{\apb@width}
\newcommand{\autoparbox}[2][c]{\settowidth{\apb@width}{#2}\parbox[#1]{\apb@width}{#2}}

\makeatother

\begin{document}
\begin{flushright}\footnotesize
\texttt{AEI-2009-083}
\end{flushright}

\begin{center}
{\Large\textbf{\mathversion{bold}%
A Supersymmetry preserving Mass-Deformation of $\mathcal{N}=1$ Super Yang-Mills in $D=2+1$}\par}
\vspace{1cm}

%\textsc{Abhishek Agarwal}
\textsc{Abhishek ~Agarwal}\\
\vspace{5mm}
\textit{American Physical Society\footnote{Present address.}\\
1 Research Road\\
Ridge, NY 11961, USA}\\
\vspace{2mm}
\textit{and}\\
\vspace{2mm}
\textit{ Max Planck Institut f\"{u}r Gravitationsphysik\\
Albert Einstein Institut\\
Am M\"{u}hlenberg-1, D14476, Potsdam, Germany}\vspace{3mm}

\verb+ abhishek@ridge.aps.org+
\par\vspace{1cm}

\vfill

\textbf{Abstract}\vspace{5mm}

\begin{minipage}{12.7cm}%\centering
We construct a massive non-abelian $\mathcal{N}=1$ SYM theory on $\mathbb{R}^3$.
This is achieved by  using a non-local gauge and Poincare invariant mass term for gluons due to Nair. The underlying supersymmetry algebra is shown to be a non-central extension of the Poincare algebra by the spacetime rotation group $so(3)$. The incorporation of Chern-Simons couplings in the formalism  is also presented. The dimensional reduction of the gauge theory and the SUSY algebra is related to a massive $\mathcal{N}=2$ massive matrix quantum mechanics based on euclidean $Clifford_2(\mathbb{R})$.

\end{minipage}

\end{center}
\vfill
\bibliographystyle{h-physrev}
\newpage
\setcounter{page}{1}
\pagenumbering{arabic}
\renewcommand{\thefootnote}{\arabic{footnote}}
\setcounter{footnote}{0}

%%%%%%%%%%%%%%%%%%%%%%%%%%%%%%%%%%%%%%%%%%%%%%%%%%%%%%%%%%%%%%%%%%%%%%%%%%%%%%%%%%%%%%%%%%%%%%%%%%%%%%%%%%%%%%%%%%%%%%%%%%%%%%%%%%%%
%%%%%%%%%%%%%%%%%%%%%%%%%%%%%%%%%%%%%%%%%%%%%%%%%%%
\section{Introduction}
Yang-Mills theories in three spacetime dimensions $(YM_{2+1})$ provide a fascinating testing ground for ideas related to non-perturbative aspects of gauge theories. For instance, pure Yang-Mills in $D=2+1$ is one of the few non-supersymmetric gauge theories that can be studied in a manifestly gauge invariant Hamiltonian formalism. Starting from first principles, the formalism due to Kim, Karabali and Nair (KKN) has led to an understanding of many non-perturbative features of the purely gluonic theory\cite{KKN-Papers,KKN-Mass}. The successes of the KKN approach
include an understanding of the mechanism for the generation of a mass-gap in the spectrum of  pure Yang-Mills theory  as well as a precise computation and prediction for its string tension; which compares remarkably well with the latest lattice data. More recently, progress has been achieved towards extending the formalism to incorporate matter fields and understand screening effects\cite{matter}, adapting the  Hamiltonian analysis to include non-trivial spacial geometries  such as $S^2$\cite{S2} along with promising schemes for precision computations of glueball masses\cite{minic} and strong coupling corrections to the string tension\cite{string2}. Among other impressive insights, and central to the concerns of the present paper, this list of successes also explains how the gluons of the three dimensional gauge theory become massive leading to a gapped spectrum.

In the purely gluonic theory  the mass-gap is related to the volume measure on the gauge invariant configuration space of the gauge theory. In a gauge invariant formalism, the gauge potentials can be replaced by a scalar (Hermitian) matrix degree of freedom $H$.  This can be interpreted as the non-abelian version of the standard dualization prescription for three dimensional electrodynamics, where the photon can be replaced by a scalar. The volume element on the space of gauge configurations $\mathcal{A}$ modulo gauge transformations $\mathcal{G}$: $d\mu(\mathcal{A}/\mathcal{G})$, can be shown to be related to the Haar measure on the space of $H$ $(d\mu(H))$\cite{KKN-Papers} as
\<
d\mu(\mathcal{A}/\mathcal{G}) = 2c_A S_{wzw}(H)d\mu(H)
\>
$S_{wzw}(H)$ is the Wess-Zumino-Witten action while $c_A$ is the quadratic adjoint Casimir for the gauge group. $S_{wzw}(H)$ effectively cuts off the volume of the physical configuration space resulting in a massive spectrum for the gauge theory. The effect of the measure and the resulting  mass-gap have been explored in depth by KKN\cite{KKN-Papers} in a Hamiltonian framework.

Though a covariant approach towards a non-perturbative reformulation  of three dimensional pure Yang-Mills theory remains to be developed, it is nonetheless possible to write down a gauge and poincare invariant
mass-term for $YM_{2+1}$ (which we subsequently denote by $S_m$) that is manifestly suitable for a path-integral analysis. The mass-term in question, introduced by Alexanian and Nair\cite{VPN-hot-3,AN}  can be interpreted as the magnetic mass for high temperature $QCD$. Indeed, finite temperature considerations were what led to its introduction in\cite{VPN-hot-3}. For related previous analyses of  the electric and magnetic sectors of finite temperature $QCD$ we shall refer to\cite{VPN-hot-others}.

Though the precise relation between the covariant mass-term suggested in\cite{AN, VPN-hot-3} and the much more elaborate Hamiltonian framework due to KKN is not completely understood there are several reasons to believe that $S_m$ is closely related to the volume measure on the gauge theory configuration space. As a case in point, it was shown in\cite{AN} that $S_m$ is dynamically generated in the purely gluonic theory. This analysis was performed by a self consistent  rearrangement of the perturbation series and the resulting (leading order) estimate for the mass-gap is strikingly close to the exact answer, which is now known from the Hamiltonian KKN framework\footnote{The leading estimate in \cite{AN} for the mass gap $m \approx 1.2 m_{exact}$}. Furthermore, an observation, which is   more compelling from the analytical point of view, was  made in\cite{KKN-Mass}. In that paper, it was shown that a particular covariantization of the mass-term obtained in the Hamiltonian framework from the volume measure led directly to $S_m$. Though covariantization does not uniquely lead to $S_m$; an alternate possibility was also discussed in\cite{AN}, it has several welcome algebraic attributes. For instance, it was shown in\cite{AA-mass-1} that $S_m$ is the unique covariantization of the configuration space volume measure that leads to standard gaussian mass-terms for the matrix quantum mechanics obtained by the dimensional reduction of the gauge theory. Furthermore, as we show in the present paper, it admits a very natural supersymmetrization. Thus, although the precise origins of $S_m$ and its potential relation to the measure on the gauge  theory configuration space remain partly obscure, it is clear that it deserves to be investigated further. In the present paper, we shall simply consider it as a potential gauge and Lorentz invariant mass-term that one can use to deform the gluonic theory and make it manifestly massive. With this particular approach in mind, we shall proceed to work out its generalization that has $\mathcal{N} =1$ supersymmetry.

Three dimensional supersymmetric gauge  theories are also of considerable interest from the point of view of gauge/gravity duality. For instance, $\mathcal{N} =8$ SYM, being related to the worldvolume theory of D2 branes is of fundamental significance  in extending and testing the gauge/gravity duality in the regime of non-conformal gauge theories. Though this particular gauge theory on $\mathbb{R}^3$ is massless, its spectrum can be rendered massive by recasting it on $\mathbb{R} \times S^2$\footnote{Recall that in three spacetime dimensions Yang-Mills theories are generically not conformal. Thus there is no natural way to map the results for the theory on $\mathbb{R} \times S^2$ to those for the theory defined on $\mathbb{R}^3$.}. A string dual for this  massive sixteen supercharge theory has recently been proposed in\cite{lm}  and several analytical features pertaining to its spectrum have also been extracted in the same paper. Apart from super Yang-Mills theories, sixteen and twelve supercharge super Chern-Simons theories have also been investigated in great deal in the recent past in the wake of the exciting proposals for their gravity duals and M2 brane dynamics\cite{blg,abjm}. The two sets of developments are not completely independent from the point of view of the duality between D2 and M2 brane theories\cite{seiberg1} as the sixteen supercharge supercharge super Chern-Simons theory is indeed expected to describe the IR dynamics of the corresponding Yang-Mills theory. For recent evidence in this regard at the Lagrangian level see\cite{mukhi}. Interestingly the super Chern-Simons theories mentioned above admit explicit massive deformation on $\mathbb{R}^3$, which have also been worked out  in\cite{SCS-mass}.

Apart from the massive nature of the spectrum, a  central feature common to the  supersymmetric gauge theories mentioned above is that the underlying supersymmetry algebra takes on the following schematic form:
\<
[\mathcal{Q}, \mathcal{Q}]_+ \sim P + mR \label{schematic}
\>
where $m$ is the characteristic mass scale and $R$ is a (flavor) R-symmetry generator. Such mass-deformed algebras
also appear as the symmetry algebra  of the scattering matrix of the spin-chain/planar dilatation operator of $\mathcal{N} =4$ SYM and the dual worldsheet theory\cite{n=4} as well as in the closely related plane wave matrix model\cite{bmn, ppwmm}. Their appearance as the spacetime symmetry algebra of three dimensional gauge theories leads to remarkable consequences. For instance, it is possible to use the algebra to completely constrain all four particle scattering amplitudes in the corresponding super Chern-Simons theories, to $all$ orders in perturbation theory up to a single undetermined function\cite{abm}. This result parallels the developments previously known result in the case of the scattering matrix of the dilatation operator of $\mathcal{N}=4$ SYM and the dual worldsheet computations\cite{n=4}.  The non-central extensions of the supersymmetry algebras are a natural consequence of the massive nature of the underlying theories in both the cases and the severe constraining power of the algebra so obtained is an  extremely welcome feature.

A key point of departure for the theories we consider in this paper from the examples cited above is that the theories in question do not have extended supersymmetry, and hence they lack
$R$  symmetry.  Nevertheless, the mass deformation of $\mathcal{N}=1$ SYM theory that we perform leads to an algebraic structure of the form (\ref{schematic}), where $R$ stands for the spacetime $so(3)$ rotations. Our construction may also be contrasted with the widely known method for rendering Yang-Mills theories massive in three dimensions; namely the addition of Chern-Simons terms, see for example\cite{witten3dindex}.   An $\mathcal{N} =1$ deformation of a Super Yang-Mills theory by Chern-Simons terms and massive Fermions does $not$ lead to a deformation of the underlying supersymmetry algebra, while for the present case it does. Thus, from the point of view of the underlying algebra, the massive gauge theories constructed   in this paper is closer in spirit to the  theories studied in \cite{lm,SCS-mass}, even though the  Yang-Mills theories in question are defined on $\mathbb{R}^3$ with minimal supersymmetry, as opposed to the case in \cite{lm} where the SYM has extended SUSY and is defined on $\mathbb{R}\times S^2$ or the examples studied in \cite{SCS-mass}, where the gauge theories do not have any Yang-Mills terms in their actions. It is also worth noting that non-central extensions of the type  (\ref{schematic}) are very constrained, as a generic lie algebra $R$ will lead to a violation of the super Jacobi identity. However, as we show later in the paper,  the three dimensional spacetime rotations $so(3)$  do lead to a consistent extension of the super-Poincare algebra in three dimensions, which in turn, makes the mass-deformations of the $\mathcal{N}=1$ theory of the kind that we explore in the paper possible.

The results presented here are closely related to the observations noted in\cite{AA-mass-1} where mass-deformations of various supersymmetric three dimensional Yang-Mills theories were related to supersymmetric matrix models by dimensional reduction. However, the issue of whether or not the  deformed Yang-Mills theories are supersymmetric themselves was not answered in that paper. In this paper we address  that question in the context of the minimally supersymmetric SYM while deferring the issue of extended supersymmetry to a future publication.

The paper is organized as follows. We begin with a brief review of the mass-term introduced in \cite{AN, VPN-hot-3} and its various relevant algebraic properties. Following that we proceed to present the supersymmetrization of the mass-deformed theory and extract the underlying supersymmetry algebra. This construction is followed by a discussion of how Chern-Simons terms may also be introduced within our framework and we work out the interplay between the mass $(m)$ used in deforming the algebra and Chern-Simons level number.  We end the paper with a discussion of how the dimensional reduction of the mass-deformed theory constructed in the paper can be related to massive $\mathcal{N}=2$ matrix quantum mechanics, the unique example of which was worked out in\cite{kimpark}. We also comment on the consistent contraction of the supersymmetry algebra that results from the dimensional reduction.
\section{Massive Bosonic $YM_{2+1}$: A Brief Review}
The action for pure Yang-Mills with the mass term included can be written as:
\[ S_{YM} =\int d^3x \frac{1}{4g^2} F^a_{\mu \nu} F^a_{\mu \nu} + \frac{1}{g^2}S_m.\label{S}\]
 $S_m$\cite{AN} is the mass term introduced in\cite{AN}.  As explained below and elaborated upon in the appendix, many of the features of this term are best elucidated by expressing it in terms 
 of seemingly two dimensional gauge potentials $A_\pm$.  It is important to stress that the apparently  two dimensional quantities $A_\pm$ etc are generated from $A_\mu$ by contracting them with a set of auxiliary three dimensional null vectors $n, \bar{n}$, and  this apparent  two dimensionality is not to be confused with the actual dimensionality of the spacetime. The null-vectors are in turn taken to depend on the coordinates of an associated $S^2$ which we refer to as $\Omega$. The spherical coordinates have nothing to do with the three dimensional spacetime, and for all practical purposes they can be regarded as a bookkeeping device. Nevertheless, this pseudo two dimensional formalism is extremely useful for many of the computations carried out later and it is utilized heavily throughout   the paper. The details about the conventions regarding the null vectors and the definitions of $A_\pm$ can be found in the appendix.  \\
$S_m$ can be expressed in  explicit detail as
\[ S_m = m^2\int dx_0 d\Omega K(A_+,A_-) \label{SM},\]
where the kernel $K$ is given by
\[
K(A_+,A_-) = -\frac{1}{\pi}\int_1\left(\tr (A_+(1)A_-(1)) + i\pi I(A_+(1)) + i\pi I(A_-(1))\right).
\]
While
\[ I(A(1)) = i\sum_n \frac{(-1)^n}{n}\int _{2\cdots n}\frac{\tr (A(1)\cdots
  A(n))}{\bar{z}_{12}\bar{z}_{23}\cdots \bar{z}_{n1}}
\frac{d^2x_1}{\pi}\cdots \frac{d^2x_n}{\pi}.\label{I}\]
The arguments of $A$ refer to the different `spacial' points. The transverse coordinate $x_0$ is the same for all
the $A$'s in the above expression for $I$.
Alternatively, the mass term can also be formally expressed as
\[
K(A_+,A_-) = -\tr\left(\frac{A_+A_-}{\pi} + \ln(D_+) + \ln(D_-)\right)\label{trlog}
\]
where $D_\pm = \partial_\pm + A_\pm$.
The trace in the above expression stands for the trace over the color indices as well as the integration over the transverse coordinates. 

If $A_\pm $ had really been the components of a two dimensional gauge potential then the formal expression above would have related $K$ to a gauged WZW model;  a
fact that is well known from studies of  $QCD$ in two dimensions \cite{2dqcd}. What we have above is a three dimensional version of the WZW functional, where the $D=2+1$ gauge potentials are organized in terms of two dimensional quantities $A_\pm$ is a somewhat twistorial fashion.

Although $S_m$ is obviously non-local, as might be expected for a mass-term for gluons, it has a perfectly well defined expansion in powers of the gauge potential making it suitable for the standard perturbative (loop) expansion. For instance, the first three terms
in the expansion of the kernel $K$ can be expressed as\cite{AN}
\<
\int dx_0 d\Omega K(A_+,A_-) = K_2 + K_3 + K_4 + \cdots
\>
\<
K_2 \eq \frac{1}{2}\int_k A_\mu^a(k)\left[\delta_{\mu\nu} - \frac{k_\mu k_\nu}{k^2}\right]A_\nu^a(-k)\nln
K_3 \eq  \int_{k_i,\Omega}  \frac{i}{12\pi }\tr\left(A(k_1).n[A(k_2).n,
  A(k_3).n]\right)\left(\frac{1}{k_1.n}\left(\frac{k_2.\bar{n}}{k_2.n}
    - \frac{k_3.\bar{n}}{k_3.n}\right)\right).\nln
K_4 \eq -\frac{1}{8\pi} \int_{k_i,\Omega} \ \frac{\tr(A.n(k_1)\cdots
A.n(k_4))}{k_3.n +
  k_4.n}\left(\frac{1}{k_2.n}\left(\frac{k_3.\bar{n}}{k_3.n}-\frac{k_4.\bar{n}}{k_4.n}\right)-
\frac{1}{k_1.n}\left(\frac{k_3.\bar{n}}{k_3.n}-\frac{k_4.\bar{n}}{k_4.n}\right)\right)\nln
\>
Overall conservation of momenta is implied in the above formulae for the vertices.

The contributions to these vertices to the gluon self-energy were computed in\cite{AN}.

In what is to follow, it would be convenient to regard the action described above as being obtained from a `Lagrangian' on $ S^2$. The dynamical degrees of freedom of the theory have no dependence on $S^2$, which, as mentioned before,  is only used as a bookkeeping device. However, it would be useful to relegate the $S^2$ integral to  the very end. The $S^2$ valued functional  which we shall investigate is:
\[
\mathcal{S} =  \int d^3x \frac{3}{2\pi g^2} F^a_{+ -} F^a_{- +} + \frac{1}{g^2}\mathcal{S}_m.\label{S2}.\]
It is implied $\mathcal{S}_m$ is the mass term given above with the $\Omega $ integrals left unevaluated. One can easily verify that
\[ \int_\Omega \mathcal{S} = S_{YM}\]
The equations of motion, without integrating over  $\Omega$ integration, can be written as:
\<
\frac{3}{\pi} (D_-F_{-+})^a + m^2 J_-^a \eq 0, \hspace{.3cm}  \frac{3}{\pi} (D_+F_{+-})^a + m^2 J_+^a = 0
\>
The currents
\<
J^a_\pm \eq \frac{1}{\pi}\tr\left(-it^a[\mathcal{A}_\pm - A_{\pm}]\right)
\>
involve the auxiliary gauge fields $\mathcal{A}_\pm$ which satisfy an associated Chern-Simons equation of motion: namely,
\<
D_+\mathcal{A}_- \eq \partial _- A_+, \hspace{.3cm}  D_-\mathcal{A}_+ = \partial _+ A_-\label{atoA}
\>
The Chern-Simons equations obviously originate from the variation of the kernel $K$ used in defining $S_m$ (\ref{SM}).  As pointed out in \cite{VPN-hot-3}, $K$, or more precisely $I$ (\ref{I}), thus has a natural interpretation  as  the eikonal of an associated Chern-Simons theory.

The equations for $\mathcal{A}_\pm$, in turn, imply
\[
D_\mp J^a_\pm = \frac{1}{2\pi}F^a_{\mp \pm}\label{J2F}
\]
Thus, though the current $J$ is highly non-local, it enjoys the special property of being related to the field strength through the action of the covariant derivative.  This particular feature may be contrasted with the case one would have encountered if a Chern-Simons term, instead of $S_m$  was chosen as the mass term. In the Chern-Simons case, the corresponding current $\tilde{J}_\mu$ (in the $\mathbb{R}^3$ notation) i.e.  the variation of the Chern-Simons term would itself have been proportional to $F_{\mu \nu}$ i.e $\tilde{J}_\mu \sim \epsilon_{\mu \nu \rho}F_{\nu \rho}$. However, in the present case, it is the covariant derivative of the current, and not the current itself that is related to the field strength.  This fact is crucial to the ensuing supersymmetrization that we present in the following section, which, unlike the case of the supersymmetric Yang-Mills-Chern-Simons theory \cite{witten3dindex}, requires  the underlying super Poincare algebra to be mass-deformed as well.
\section{$\mathcal{N} =1$ SYM and its Mass-Deformation}
The action for standard massless $\mathcal{N} =1$, $D=3$ euclidian SYM is given by
\<
S \eq \frac{1}{g^2} \int_{R^3}\left(  \frac{1}{4} F^a_{\mu \nu}F^a_{\mu \nu} + \frac{1}{2}\bar{\Psi}^a \sigma_\mu D_\mu \Psi^a\right) = \frac{1}{g^2} \int_{R^3,\Omega}\left( \frac{3}{2\pi}F^a_{+-} F^a_{-+} + \frac{3}{8\pi}\bar{\Psi}^ah_{MN}D_M\sigma_N \Psi^a\right)\nln
\>
$h$ is a `world-sheet' metric
\[
h = \matr{cc}{0&+\\+&0}
\]
The massless $\mathcal{N}=1$ action is invariant under the transformations
\<
\delta_\alpha A^a_\mu \eq \bar{\alpha}\sigma_\mu \Psi^a, \hspace{1cm} \delta_\alpha \Psi^a = -\frac{i}{2}\varepsilon_{\mu \nu \rho} F^a_{\mu \nu}\sigma_\rho \alpha \nln
\delta_\alpha A^a_\pm \eq \bar{\alpha} \sigma_\pm \Psi^a, \hspace{1cm} \delta_\alpha \Psi^a = \left(6[\sigma_+,\sigma_-]F_{+ -}\right)\alpha
\>
The two ways of expressing the supersymmetry transformations are related, as the second form implies the former upon an $S^2$ integration of both sides of the equation.  In the second form, the auxiliary $S^2$ coordinates are not integrated out and working with it will simplify the task of constructing the mass deformation of the $\mathcal{N} =1$ transformations. Notice that in the two dimensional formalism the fermions do not depend on the $S^2$ coordinates, even though the gauge fields $A_\pm$ do.

The fermions  obey  a majorana condition $\bar{\Psi} = \Psi ^t\epsilon$ where the charge conjugation matrix; $\epsilon = -i\sigma_2$, satisfies $\sigma_\mu^t \epsilon = -\epsilon \sigma _\mu$.

For the mass deformed case, we make the following ansatz for the action $S = \frac{1}{g^2}\int_\Omega \mathbb{S}$, and the supersymmetry transformations.
\<
 \mathbb{S} \eq \int_{R^3}\left[ \frac{3}{2\pi}F^a_{+-} F^a_{-+} + \mathcal{S}_m + \frac{3}{8\pi}\bar{\Psi}^ah_{MN}D_M\sigma_N \Psi^a + \frac{\omega m}{8\pi}\bar{\Psi}^a\Psi^a\right] \label{n=1}
\>
\<
\delta_\alpha A^a_\pm \eq \bar{\alpha} \sigma_\pm \Psi^a\nln
\delta_\alpha \Psi^a \eq (\delta^0_\alpha + \delta^1_\alpha) \Psi^a = \left(6[\sigma_+,\sigma_-]F_{+ -} + \gamma  m (h_{AB}\sigma_AJ_B)\right)\alpha \nln
\eq \left(3[\sigma_A,\sigma_B]F_{CD}h_{AD}h_{BC} + \gamma  m (h_{AB}\sigma_AJ_B)\right)\alpha \label{msusy}
\>
$\omega$ and $\gamma $ are numerical factors that are to be determined from the condition $\delta_\alpha \mathbb{S} = 0$. $\delta^0, \delta^1$ generate the terms of $\mathcal{O}(1),  \mathcal{O}(m)$ on the $r.h.s$ respectively.

Since the action as well as the supersymmetry transformations reduce to the standard $\mathcal{N}=1$ case in the massless limit, it is follows that
\[
\delta_{\alpha}\mathbb{S}|_{m=0} = 0
\]
To examine the $\mathcal{O}(m)$ terms in the supersymmetry variation of the action, we note that:
\<
\delta^1_\alpha \int  \frac{3}{8\pi}\bar{\Psi}^ah_{IJ}D_I\sigma_J \Psi^a
\eq \frac{3m\gamma }{8\pi}\int h_{AB}h_{IJ}[D_IJ_B]\left[\bar{\Psi} ([\sigma_J,\sigma_A]_+ +[\sigma_J,\sigma_A])\alpha\right]
\>
Using $h_{AB}D_AJ_B = 0$ and (\ref{J2F}) we have
\[
\delta^1_\alpha \int  \frac{3}{8\pi}\bar{\Psi}^ah_{IJ}D_I\sigma_J \Psi^a = -\frac{3}{8\pi}\frac{2m\gamma}{2\pi}\int (\bar{\Psi} ^a[\sigma_+,\sigma_-]\alpha)F^a_{+-}\label{del11}
\]
In the calculations  leading up to this, we have suppressed the color superscipts, to avoid confusion with the spinor indices.

We also note that
\[
\frac{\omega m}{8\pi}\delta ^0_{\alpha} \int \bar{\Psi}\Psi = 2\frac{6\omega m}{8\pi} (\bar{\Psi} ^a[\sigma_+,\sigma_-]\alpha)F^a_{+-}\label{del12}
\]
From (\ref{del11}) and (\ref{del12})  we hence get:
\[
\frac{\partial \delta _\alpha \mathbb{S}}{\partial m}|_{m=0} = 0 \hspace{.4cm} \Rightarrow   \hspace{.4cm} \gamma  = 4\pi\omega
\]
To analyze the $\mathcal{O}(m^2)$ terms, we note that:
\[
\delta_\alpha \mathcal{S}_m = m^2\int h_{AB}J_A(\bar{\alpha }\sigma_B \Psi), \hspace{.4cm} \delta^1_\alpha \left[\frac{\omega m}{8\pi} \int \bar{\Psi}\Psi\right] = -\frac{\gamma \omega m^2}{4\pi}\int h_{AB}J_A(\bar{\alpha }\sigma_B \Psi)
\]
Thus:
\[
\frac{\partial ^2\delta _\alpha \mathbb{S}}{\partial m^2}|_{m=0} = 0 \hspace{.4cm}\Rightarrow  \hspace{.4cm} \gamma \omega = 4\pi
\]
Thus a consistent solution for the mass-deformed ansatz is given by:
\[
\omega =1, \hspace{1cm} \gamma = 4\pi
\]
For these value of the parameters the action (\ref{n=1}) is invariant under the supersymmetry transformations (\ref{msusy}).\\
Having obtained the supersymmetry transformations in the two dimensional notation, it is also instructive to depict their form in the explicit $\mathbb{R}^3$ form.
Integrating both sides of the supersymmetry transformations (\ref{msusy}) over $S^2$ we can express them as follows:
\<
\delta_\alpha A^a_\mu \eq \bar{\alpha} \sigma_\mu \Psi^a\nln
\delta_\alpha \Psi^a \eq (\delta^0_\alpha + \delta^1_\alpha) \Psi^a =  -\frac{i}{2}\varepsilon_{\mu \nu \rho} F^a_{\mu \nu}\sigma_\rho \alpha +   m J^a_\mu\sigma_\mu\alpha\label{msusyr3}
\>
where,
\[
J^a_\mu = \frac{1}{2}\int_\Omega (J^a_+\bar{n}_\mu + J^a_-n_\mu)
\]
We also note that, in the abelian case,
\<
J_\mu \eq (A_\mu - \partial_\mu \frac{1}{\partial^2}(\partial_\nu A_\nu))\nln
\delta_\alpha \Psi \eq  -\frac{i}{2}\varepsilon_{\mu \nu \rho} F_{\mu \nu}\sigma_\rho \alpha +   m (A_\mu - \partial_\mu \frac{1}{\partial^2}(\partial_\nu A_\nu))\sigma_\mu\alpha
\>
As a further consistency check, the abelian action  expressed  in manifestly $\mathbb{R}^3$ notation as;
\<
S \eq \frac{1}{g^2}\int_{\mathbb{R}^3}\left( \frac{1}{4}F_{\mu \nu}F_{\mu \nu} + \frac{m^2}{2}A_\mu(\delta_{\mu \nu}-\partial_\mu \frac{1}{\partial^2}\partial_\nu)A_\nu + \frac{1}{2}\bar{\Psi}(\sigma_\mu \partial_\mu + m)\Psi \right)\label{sabelian}
\>
is readily checked to be invariant under the $\mathbb{R}^3$ form of the SUSY transformations (\ref{msusyr3}).

To verify the invariance  of the non-Abelian  action under (\ref{msusyr3}), one needs the $\mathbb{R}^3$ version of the relation (\ref{J2F}); namely:
\[
D_{[\mu}J^a_{\nu]} = F^a_{\mu \nu}. \label{j2fr3}
\]
This relation is crucial for the $\mathcal{O}(m)$ terms to vanish in the SUSY variation of the action. It is trivially verified in the Abelian case using the explicit form of the current $J_\mu$ given above. In the non-Abelian case, starting from the equations (\ref{J2F}) and taking their difference, we get (we suppress the color indices in what follows):
\[
D_\mu(J_- n_\mu - J_+\bar{n}_\mu ) = \frac{i}{2\pi}\epsilon_{\alpha \beta \gamma }F_{\alpha \beta }x_\gamma
\]
Multiplying this equation by $x_\rho$ and integrating over $\Omega $ gives
\<
\frac{2i}{3}\epsilon_{\alpha \beta \rho } F_{\alpha \beta } \eq D_\mu \int_\Omega (J_- n_\mu - J_+\bar{n}_\mu )x_\rho \nonumber\\
\eq \frac{4i}{3} \epsilon _{\mu \nu \rho}\partial_\mu (A_\nu - \partial_\nu \frac{1}{\partial^2}(\partial_\lambda A_\lambda)) + \cdots\label{j2fr3proof}
\>
The second line gives   the terms linear in the gauge potential $A$, which agrees with the abelian limit. Since the integrand on the $l.h.s$ involves the non-abelian completion of $J$, we can argue based on general gauge covariance that
\[
D_\mu \int_\Omega (J_- n_\mu - J_+\bar{n}_\mu )x_\rho = \frac{4i}{3} \epsilon _{\mu \nu \rho}D_\mu J_\nu \hspace{.2cm}\Rightarrow \hspace{.2cm} D_{[\mu}J^a_{\nu]} = F^a_{\mu \nu}
\]
To phrase the argument slightly differently: The gauge covariance of the left and right hand sides of the first line in (\ref{j2fr3proof}) allow us to make the ansatz:
\[
D_\mu \int_\Omega (J_- n_\mu - J_+\bar{n}_\mu )x_\rho  = \beta \epsilon _{\mu \nu \rho}D_\mu J_\nu
\]
where $\beta $ is to be determined. The form of the $r.h.s$ above is validated by taking the Abelian limits of both sides of the equation above. Once the form is  fixed, the  undetermined constant can also be determined by evaluating the $l.h.s $ explicitly in  the Abelian limit.\\
Thus to summarize the main results derived above; we have found a mass deformation of $\mathcal{N}$ =1 SYM theory in three dimensions (\ref{n=1}) which is invariant under the supersymmetry transformations (\ref{msusy}) (or equivalently (\ref{msusyr3})).
\subsection{Determination of the Algebra}
In this section, we shall try and derive the superalgebra underlying the supersymmetrization carried out above by computing the commutator of SUSY variations on the gauge potential. For the purposes of extracting the algebra from the closure on $A_\mu$ it suffices to consider the abelian limit, where matters simplify significantly.\\
We start the analysis with the abelian case in the $\mathbb{R}^3$ notation.  It is easily seen that
\<
\delta_{[\rho}\delta_{\omega]}A_\mu \eq -\frac{i}{2}\epsilon_{\alpha \beta \gamma}F_{\alpha \beta}(\bar{\omega}[\sigma_\mu, \sigma_\gamma]\rho) + mJ_\nu ((\bar{\omega}[\sigma_\mu, \sigma_\nu]\rho)\nln
\eq 2\left[-F_{\mu \lambda}(\bar{\omega}\sigma_\lambda \rho) + i m \epsilon_{\mu \nu \lambda}(\bar{\omega}\sigma_\lambda \rho)J_\nu\right]\nln
\eq \partial_\mu (-2\vec{A}.(\bar{\omega}\vec{\sigma}\rho) + 2(\bar{\omega}\vec{\sigma}\rho).\vec{\partial}A_\mu  + 2 i m \epsilon_{\mu \nu \lambda}(\bar{\omega}\sigma_\lambda \rho)A_\nu - 2 i m  \epsilon_{\mu \nu \lambda}(\bar{\omega}\sigma_\lambda \rho)\left[\partial _\nu \frac{1}{\partial^2}\vec{\partial}.\vec{A}\right]\nln \label{abelian-double-comm-A}
\>
The first term in the last line is a gauge transformation. The second and third terms are translations and rotations of the gauge potential while the last term is a gauge transformation followed by a rotation. It is worth noting that the gauge parameter in the last term is a non-local quantity, $\frac{1}{\partial^2}\vec{\partial}.\vec{A}$, which is a manifestation of the non-local nature of the mass-term. Thus, discarding the terms involving gauge transformations, we have:
\<
\delta_{[\rho}\delta_{\omega]}A_\mu \eq 2(\bar{\omega}\vec{\sigma}\rho).\vec{\partial}A_\mu  + 2 i m \epsilon_{\mu \nu \lambda}(\bar{\omega}\sigma_\lambda \rho)A_\nu\nln
\eq 2(\bar{\omega}\sigma_\nu\rho)(iP_\nu A_\mu - i m \epsilon_{\mu \nu \beta}A_\beta).
\>
We see that the supersymmetry algebra underlying the mass deformed $\mathcal{N}=1$ theory corresponds to a non-central extension of the $\mathcal{N} =1$ algebra. The anti-commutator of supercharges closes on translations and $so(3)$ spacetime rotations.
\<
[\bar{\beta}\mathcal{Q}, \bar{\alpha}\mathcal{Q}] \eq 2i(P_a - m \mathcal{R}_a), \hspace{1cm} \vec{a} = \bar{\alpha }\vec{\sigma} \beta \label{msusy-alg}
\>
It is understood that
\<
[P_aA]_\mu \eq \vec{a}.\vec{p}A_\mu, \hspace{.5cm} [\mathcal{R}_aA]_\mu = \epsilon_{\mu \nu \rho} a_\nu A_\rho, \hspace{.5cm} [\mathcal{R}_a\Psi]_b =  (\vec{a}.\vec{\sigma})_{bm}\Psi_m
\>
It is also important to verify that the Jacobi identity
\<
[\bar{\gamma}\mathcal{Q}, [\bar{\beta}\mathcal{Q},\bar{\alpha}\mathcal{Q}]]+ [\bar{\beta}\mathcal{Q},[\bar{\alpha}\mathcal{Q},\bar{\gamma}\mathcal{Q}]]+ [\bar{\alpha}\mathcal{Q},[\bar{\gamma}\mathcal{Q},\bar{\beta}\mathcal{Q}]] \eq 0
\>
is satisfied. The non-trivial part of the identity involves the commutators of $\mathcal{R}$ with $\mathcal{Q}$, which translates into the requirement that:
\<
(\bar{\beta}\vec{\sigma}\alpha).(\bar{\gamma}\vec{\sigma}\mathcal{Q}) + (\bar{\alpha}\vec{\sigma}\gamma).(\bar{\beta}\vec{\sigma}\mathcal{Q})+
(\bar{\gamma}\vec{\sigma}\beta).(\bar{\alpha}\vec{\sigma}\mathcal{Q}) \eq 0
\>
A straightforward computation can be used to verify that this is indeed satisfied for arbitrary spinors $\alpha, \beta $ and $\gamma$.
Thus the $\mathcal{N}=1$ algebra underlying the mass deformed model is nothing but a non-central extension of $osp(1|2)$. The extension in question being brought about by the spacetime rotation group $so(3)$.

As noted in\cite{abm}, the appearance of the mass in the algebra itself implies that it plays the role of a structure constant. Consequently, it is protected against `running' in the renormalization group sense. Thus $m$ can be regarded as a parameter, even though it is obviously a dimension-full quantity.

It is also instructive to analyze the commutator of SUSY variations in the two dimensional notation, without integrating out the $S^2$ dependence.  For example, it is easily shown in the non-abelian case  that:
\<
\delta_{[\beta,\alpha]}A^a_+ \eq -12(\bar{\alpha }\sigma_B\beta ) F^a_{+A}h_{BA} + 4\pi m(\bar{\alpha }[\sigma_+,\sigma_A]\beta ) J^a_Bh_{AB}.\label{comms2}
\>
Projecting out the $\mathbb{R}^3$ components of the above equation would once again yield a linear combination of rotation, translations, gauge transformation and gauge transformations followed by rotations. However, the specific numerical factors multiplying these transformations obtained by projecting the above equation would differ from the action of the double commutator of (\ref{msusyr3}) on $A_\mu$.  For instance, even in the massless case, it is seen that
\<
\frac{3}{16\pi} \int_\Omega (\delta _{[\beta,\alpha]})A^a_+\bar{n}_\lambda  \eq -\frac{i}{2}\frac{3}{2}(\bar{\alpha}[\sigma_\lambda , \sigma_\gamma]\beta)\epsilon_{\mu \nu \gamma}F^a_{\mu \nu}
\>
which differs from the expected answer by a factor of $3/2$. \\
The discrepancy has to do with the fact that the $S^2$ integration is to be thought of as averaging over all Lorentz transformations\cite{VPN-hot-3}, and the average of the product of two SUSY variations is obviously $not$ the same as the product of the average.  In other words, the average of the commutator yields a different linear combination of the supercharges than the one obtained by evaluating the commutator of (\ref{msusyr3}).  However, the fact that the $r.h.s$ of  (\ref{comms2}) would close  on translations and rotations (modulo gauge transformation) is guaranteed as we have already checked the action (in the $\mathbb{R}^3$ notation ) to be explicitly invariant under the transformations  (\ref{msusyr3}), which are implied by (\ref{msusy}). Furthermore,  we have shown the massive SUSY variations  to generate the algebra (\ref{msusy-alg}), which can be extracted from the abelian limit of the theory.

\section{Chern-Simons Terms}
One can  add Chern-Simon terms to the mass-deformed $\mathcal{N}=1$ SYM. In $\mathbb{R}^3$ notation, the action
\<
g^2 S = \int_{\mathbb{R}^3}\frac{1}{4}F^a_{\mu \nu}F^a_{\mu \nu} + S_{\sqrt{m(\frac{kg^2}{4\pi}+m)}} - \frac{ikg^2}{4\pi}\epsilon_{\mu \nu\rho}\int_{\mathbb{R}^3}(F^a_{\mu \nu}A^a_\rho - \frac{1}{3}f^{abc}A^a_\mu A^a_\nu A^a_\rho) +\nln \frac{1}{2}\int_{\mathbb{R}^3}\bar{\Psi}^a\left[\sigma.D + (\frac{kg^2}{4\pi}+m)\right]\Psi^a\hspace{7cm}
\>
can be verified  to be invariant under (\ref{msusyr3}).

$S_{\sqrt{m(\frac{k}{4\pi}+m)}}$ stands for the non-local gluonic mass term, with the coefficient $m(\frac{kg^2}{4\pi}+m)$ instead of $m^2$. Namely, while the fermionic mass $m_f$ is shifted by the level number of the Chern-Simons term, the bosonic mass $m_b$ is a geometric mean of $m$:  the parameter that appearns in the SUSY algebra, and $m_f$. As in the $k=0$ case, $m = m_b^2/m_f$ is the quantity that is effectively protected against renormalization, as it plays the role of a structure constant.

It is important to note that the addition of Chern-Simons terms is particularly important from the point of view of the $m \rightarrow 0$ limit. Although sufficient for our formal purposes of understanding $\mathcal{N}=1$ SUSY, the massless theory, without the Chern-Simons terms suffers from the well known problem of a parity anomaly\cite{witten3dindex}. Lattice data pointing to a trivial partition function for this pathological model has also been presented in\cite{EM}.  However, the $\mathcal{N} =1$ Yang-Mills-Chern-Simons system is a perfectly well defined theory. Thus, for the present  purposes, it is imperative that the mass-deformation that we consider be compatible with the addition of Chern-Simons terms so that we have  a consistent quantum field theory. Fortunately, as we show above, this is eminently possible.

We also note in passing that the addition of the Chern-Simons terms could have been carried out in the $S^2$ notation as well. For that purpose, it is useful to note that
\<
\frac{ik}{4\pi}\epsilon_{\mu \nu\rho}(F^a_{\mu \nu}A^a_\rho - \frac{1}{3}f^{abc}A^a_\mu A^a_\nu A^a_\rho) = \frac{k}{16\pi^2}\int_\Omega (F^a_{+-}A^a - \frac{1}{3} f^{abc}A^a_+ A^a_- A^a)
\>
which allows one to recast the Chern-Simons terms in an $S^2$ notation. It is implied that $A = A_\mu x_\mu$, which is the contraction of the Hodge dual of $A_\mu $ with $n$ and $\bar{n}$.
\section{Dimensional Reduction and Matrix Models}
In this final section, we relate the dimensional reduction of the massive $\mathcal{N}=1$ SYM theory constructed in this paper to the only known example of $\mathcal{N}=2$ massive matrix quantum mechanics, which was reported in\cite{kimpark}.
The dimensional reduction of pure Yang-Mills theory with the mass term added was worked out in a previous paper\cite{AA-mass-1}. The important insight was to notice that the defining equations for the auxiliary fields $\mathcal{A}_\pm$ can be readily solved upon the truncation of the theory to $D = 1$. The solution to (\ref{atoA}) can be written for the dimensionally reduced theory as
\<
\mathcal{A}_+ = \frac{n_0}{\bar{n}_0}A_-, \hspace{.3cm} \mathcal{A}_- = \frac{\bar{n}_0}{n_0}A_+
\>
Using this expression we have
\<
(S_m)_{0+1} =
-m^2\int d^3x d\Omega\tr\left[
   \frac{A_+A_-}{\pi} - \frac{1}{2\pi} \frac{k.\bar{n}}{k.n}
   \tr(A_+A_+) - \frac{1}{2\pi} \frac{k.n}{k.\bar{n}}
   \tr(A_-A_-)\right],
\>
where the `momentum' $k = (1,0,0)$. After evaluating the angular integrals one has
\<
(S_m)_{0+1} = -\frac{m^2V_{M^2}}{2}\int dx_0 \tr\left[
   A_j\left(\delta _{jl} - \frac{k_jk_l}{k^2}\right)A_l\right] = -\frac{m^2V_{M^2}}{2}\int dx_0 \tr\left[\sum_{l=1,2}
   A_lA_l\right]
   \>
 where $V_{M^2}/2$ is the volume of $T^2$ on which the spacial compactification is carried out. Thus as far as the pure `glue' part of the theory is concerned,
 \<
 \int d^3x \frac{1}{4g^2} F^a_{\mu \nu} F^a_{\mu \nu} + \frac{1}{g^2}S_m \stackrel{0+1}{\rightarrow} \int dx_0 \frac{1}{g^2_M}\tr\left(\frac{1}{2}(\mathcal{D}_t\Phi_i
  \mathcal{D}_t\Phi_i + m^2 \Phi_i\Phi_i) - \frac{1
    }{4}[\Phi_i,\Phi_j]^2\right)
\>
The matrix model coupling
\<
g^2_M = \frac{g^2}{V_{M^2}}
\>
while the hermitian matrix degrees of freedom
\<
\Phi _l = iA_l, \hspace{.3cm}\mbox{l=1,2}.
\>
Thus the dimensional reduction of the mass-deformed gauge theory is nothing but the standard mass deformation of a gauged matrix quantum mechanics of two Hermitian matrices.

In an analogous fashion, the dimensional reduction of the mass-deformed $\mathcal{N} =1$ theory produces the following matrix model action $\tilde{S}$:
\<
\tilde{S} = \int dx_0 \frac{1}{g^2_M}\tr\left(\frac{1}{2}(\mathcal{D}_t\Phi_i
  \mathcal{D}_t\Phi_i + m^2 \Phi_i\Phi_i) - \frac{1
    }{4}[\Phi_i,\Phi_j]^2 + \frac{1}{2}\bar{\Psi}( \mathcal{D}_0 + m)\Psi - \frac{i}{2}\bar{\Psi}\sigma ^i[\Phi^i,\Psi]\right)\nln\label{matno}
\>
It also follows that the current $J_\mu$ is proportional to  $(0,\Phi_1,\Phi_2)$ upon the dimensional reduction. Importantly, the resultant matrix model is $not $ invariant under the dimensional reduction of the supersymmetry transformations (\ref{msusyr3}). The supersymmetry transformations on $\mathbb{R}^3$  fail to remain a symmetry of the dimensionally reduced model
as dimensional reduction on $T^2$ breaks the  Poincare invariance of the gauge theory. Thus relations such as $D_{[\mu}J_{\nu]} = F_{\mu \nu}$, which were crucial in establishing the supersymmetry of the gauge theory on $\mathbb{R}^3$, fail to hold upon dimensional reduction. However, the lack of Poincare invariance, can be compensated for by introducing an asymetry between the bosonic and fermionic masses and making the supersymmetry transformation time dependent. The resulting supersymmetric $\mathcal{N} =2$ matrix model action is given by:
\<
\tilde{S} = \int dx_0 \frac{1}{g^2_M}\tr\left(\frac{1}{2}(\mathcal{D}_t\Phi_i
  \mathcal{D}_t\Phi_i + m^2 \Phi_i\Phi_i) - \frac{1
    }{4}[\Phi_i,\Phi_j]^2 + \frac{1}{2}\bar{\Psi}( \mathcal{D}_0 + \frac{3}{2}m)\Psi - \frac{i}{2}\bar{\Psi}\sigma ^i[\Phi_i,\Psi]\right)\nln\label{matn1}
\>
This action is invariant under
\<
\delta \Phi_i \eq \bar{\alpha}(t) \sigma^i \Psi,\nln
\delta \Psi \eq (-\sigma^{ti}\mathcal{D}_t\Phi_i + i[\Phi_i,\Phi_2]\sigma^{12} + m \sigma ^i\Phi_i)\alpha(t),\nln
\alpha(t) \eq e^{\frac{1}{2}m\sigma^0t}\alpha_0,
\>
where $\alpha_0$ is a constant spinor.

The mass of the supersymmetric matrix model (\ref{matn1})differs from that of the dimensional reduction of the $\mathcal{N} =1$ gauge theory (\ref{matno}). Thus the supersymmetric matrix model can be  thought of  as the reduction of the $\mathcal{N}=1$ theory followed by a time dependent field redefinition of the fermions. At the same time, the asymmetry introduced between the three spacetime dimensions by compactifying two of them has to be compensated for by making the supersymmetry transformation time-dependent.

Evaluating the double commutators on $\Phi_i$, we have
\<
\delta_{[\beta}\delta_{\alpha ]} \Phi_i \eq 2(\bar{\alpha}\sigma^0\beta)(\partial_t\Phi_i -im\epsilon_{ij}\Phi_j)
\>
In other words
\<
[\bar{\beta}\mathcal{Q}, \bar{\alpha}\mathcal{Q}] \eq 2i(\bar{\alpha}\sigma^0\beta)(H - m \mathcal{R}_{12}), \hspace{.5cm}\mbox{where}, \hspace{.5cm} [\mathcal{R}_{12}\Phi]_i = \epsilon_{ij}\Phi_j
\>
Comparing with (\ref{msusy-alg}) shows that the susy algebra underlying the matrix quantum mechanics is the contraction of  (\ref{msusy-alg}) to the case where there is only one spacial direction and $so(3)$ spacetime rotations are contracted to a $so(2)$ rotation between the two matrices. The  superalgebra is thus contracted from a non-central extension of $osp(1|2)$ to euclidean $Clifford_2(\mathbb{R})$.
\section{Summary and Outlook}
In the present paper we have constructed a mass-deformation of $\mathcal{N}=1$ SYM (with or without Chern-Simons couplings) based on a con-central extension of the three dimensional super-Poincare algebra by the spacetime rotation group $so(3)$. Furthermore, a consistent dimensional reduction of the gauge theory as well as the underlying supersymmetry algebra has been shown to be related to the unique example of massive $\mathcal{N} =2$ matrix quantum mechanics, which was obtained independently in\cite{kimpark} by a mass-deformation of the dimensional reduction of standard $\mathcal{N}=1$ SYM in three dimensions. The construction presented in the paper opens up the possibility of several intriguing lines of investigation, which we briefly discuss below.

The mechanism for mass-deformation presented here can obviously be used in conjunction with or as an alternative to, the better known mechanism for making three dimensional gauge theories massive, i.e  the addition of Chern-Simons terms. A plethora of extremely important results related to spontaneous supersymmetry breaking and associated physical effects have already been obtained for the $\mathcal{N}=1$ Yang-Mills-Chern-Simons system in\cite{witten3dindex}. However, we expect the physical manifestations of $S_m$ to be fundamentally different from those of Chern-Simons couplings. For instance, in the case of pure Yang-Mills theory in three dimensions, which is known to confine, the addition of Chern-Simons couplings dramatically alters the IR behavior of the theory; leading to screening rather than confinement \cite{KKNCS}.  On the other hand $S_m$ considered in the paper $is $ a covariantization of the volume measure on the configuration space of pure Yang-Mills \cite{KKN-Mass}; which has been shown to provide a first principles explanation for  confinement and spontaneous mass generation for the purely gluonic theory\cite{KKN-Papers}. Thus it is very conceivable that the massive $\mathcal{N}=1$ theory presented here would have various new features which  would doubtless be interesting to investigate along the lines specified in\cite{witten3dindex}.

It may be possible to use the constraining power of the mass-deformed supersymmetry algebra to gain insight into various physical process of interest. For instance, following the way
 mass deformed algebras of the $su(2|2)$ were employed in \cite{abm} to constrain $all$ four particle scattering processes in a large class of mass-deformed supersymmetric Chern-Simons theories up to a single undetermined function, it is conceivable that the underlying algebra can be utilized  to constrain the form of physical quantities such as, scattering amplitudes and glueball spectra, for the massive $\mathcal{N}=1$ SYM theory discussed here.

As mentioned before tell-tale signs of a potential connection between $S_m$ and the volume measure on the gauge invariant configuration space of pure Yang-Mills theory are already known to exist\cite{AN, KKN-Mass}.  It would of course be of great interest if the supersymmetrization of $S_m$ presented here can be utilized to shed some light on the nature of the configuration space for supersymmetric Yang-Mills theories in three dimensions. On a related note, we point out that although a gauge invariant Hamiltonian framework for coupling matter fields to Yang-Mills theories in three dimensions already exists\cite{matter}, the  contribution of the matter fields to the configuration space volume remains to be understood. Perhaps understanding how the supersymmetrization of $S_m$ relates to the relevant volume element can give a controlled way of broaching this interesting open issue.

Other than the issues discussed above, most of which relate to $\mathcal{N}=1$ theories, it would be extremely interesting to analyze whether three dimensional Yang-Mills theories with extended supersymmetries can be mass-deformed in a way  that is analogous to the one discussed in this paper.\\
\\
{\bf Acknowledgements:} We are grateful to Niklas Beisert, Ansar Fayyazuddin, Tristan McLoughlin  and  Parmeswaran Nair for several illuminating discussions. We are also indebted to Niklas Beisert   for his comments on a previous version of the manuscript. Some of the preliminary work leading to the results reported in this paper was carried out at the Max Planck Institute for Gravitational Physics (Potsdam, Germany), where the author held a previous position.
\section{Appendix}
We work in three dimensional Euclidean spacetime. The vectors and integrals associated with the auxiliary $S^2$ are chosen to obey the following conventions:

The $S^2$ valued complex null vectors are taken to be:
\[
n_\mu = (-\cos(\theta) \cos(\phi) -i\sin(\phi),  -\cos (\theta) \sin(\phi) +i\cos (\phi), \sin(\theta))
\]
\[
n^2 = \bar{n}^2 = 0, \hspace{.3cm} n.\bar{n} = 2, \hspace{.3cm} \epsilon_{\mu \nu \rho} n_\nu \bar{n}_\rho = 2ix_\mu
\]
where
\[ x_1 = \sin\theta \cos\phi, \hspace{.2cm} x_2  = \sin\theta \sin\phi, \hspace{.2cm} x_3 = \cos\theta\]
On the sphere $\Omega $ of volume $4\pi $
\<
\int_\Omega n_\mu \bar{n}_\nu \eq \frac{8\pi}{3}\delta_{\mu \nu}, \hspace{.3cm}
\int_\Omega \frac{k.\bar{n}}{k.n} n_\mu n_\nu = \frac{8\pi}{3}\left[ \frac{3}{2}\frac{k_\mu k_\nu}{k^2} - \frac{1}{2} \delta_{\mu \nu}\right],\hspace{.3cm}\int_\Omega x_\mu x_\nu = \frac{4\pi}{3}\delta_{\mu \nu}
\>
\< A_+ \eq \frac{1}{2} A.n, \hspace{.2cm} A_- = \frac{1}{2} A.\bar{n},\hspace{.3cm}
D_+ = \frac{1}{2} D.n, \hspace{.2cm} D_- = \frac{1}{2} D.\bar{n},\hspace{.3cm}
\sigma _+ = \frac{1}{2} \sigma .n, \hspace{.2cm} \sigma_- = \frac{1}{2} \sigma.\bar{n}\nln \>
The sigma matrices satisfy the following relations.
\<
\sigma _+^2 \eq \sigma _-^2 = 0, \hspace{.3cm} [\sigma_+, \sigma_-]_+ = 1, \hspace{.3cm} [\sigma_-, \sigma_+] = x_\mu \sigma_\mu
\>
The skew-Hermitian gauge potentials $A_\mu = -it^a A^a_\mu$ are normalized so that $tr(t^at^b) = \frac{1}{2}\delta ^{ab}$.

\end{document}